\documentclass[a4paper,11pt]{article}
\pdfoutput=1
\usepackage{jcappub}
\usepackage[T1]{fontenc}
\usepackage{amsfonts,amsmath,amssymb}
\usepackage{graphicx}
\usepackage{color}
\usepackage{natbib}
\newcommand{\rthis}[1]{\textcolor{black}{#1}}

\newcommand{\eot}{E{\"o}t-Wash }
\title{Model Comparison tests of modified gravity from the  E{\"o}t-Wash experiment}
\author[a]{Aditi Krishak}
\author[b,1]{and Shantanu Desai,\note{Corresponding author.}}

\affiliation[a] {Department of Physics, Indian Institute of Science Education and Research, Bhopal, Madhya Pradesh 462066, India}

\affiliation[b]{Department of Physics, Indian Institute of Technology, Hyderabad, Telangana-502285, India}

\emailAdd{aditi16@iiserb.ac.in}
\emailAdd{shntn05@gmail.com}

\abstract{
Perivolaropoulos et al~\cite{Periv19} (P19) have argued that  the residual torque data in the \eot experiment \rthis{is consistent with} an oscillating signal. This  could  either be  a signature of non-local modified gravity theories \rthis{or some other systematic error in the data.}
We independently assess the viability of such an  oscillating signal in the same data using Bayesian and information theoretical criterion, to complement the frequentist analysis in P19.  We fit this data to three different parametrizations (an offset Newtonian, Yukawa, and an oscillating model), and assess the significance of the oscillating model using AIC, BIC, WAIC,  and Bayes factor. All these techniques provide decisive evidence for the oscillating model compared to the Newtonian model, \rthis{provided the phase is fixed at the same value as P19. If the phase is allowed to vary, then significance from BIC, WAIC, and Bayes based tests reduces to strong evidence, whereas only AIC still shows decisive evidence.}   Our analysis codes have been made publicly available.}
\begin{document}
\maketitle
\flushbottom
\section{Introduction}
Despite being more than a century old,  General relativity (GR) currently agrees with all  tests using solar system, binary pulsar, and gravitational wave based observations~\cite{Will, Turyshev,Stairs,Yunes16,LIGOGR,LIGOGW170817,Kahya,Boran}.
In spite of this, a large number of modified theories of gravity have been explored since the inception of GR. Most  of the recent resurgence in  these alternate theories of  gravity has been driven by the need to  address problems in Cosmology such as Dark matter, Dark Energy, Inflation,  Baryogenesis, and data driven cosmological tensions~\cite{Will93,Ferreira,Koyama,Khoury,Woodard06,Schmidt,Martin,Yunes,Poplawski,Bamba,Smoot,Ferreira19,Ishak,Desaimg}. Apart from this, a  number of alternatives have also been proposed  to resolve conceptual issues, such as the Big-Bang singularity~\cite{Hawking}, arrow of time~\cite{Ellis,Padmanabhan}, or  the quantization of gravity~\cite{Carlip,Ashtekar,Woodard09}. 

One of the most stringent probes of gravity at very short distance scales are torsion balance experiments. For over three decades, the \eot group at the University of Washington~\footnote{\url{https://www.npl.washington.edu/eotwash/node/1}} has been conducting a series of such high precision torsion balance based  tests of gravity at sub-millimeter scales   to look for departures from Newtonian gravity. Their own analysis has not revealed any signs of new physics. However, other authors~\cite{Periv19} (P19 hereafter) have independently analyzed the data from  these experiments~\cite{eot1,eot2,eot3,eot4},  and have argued  that the  residual data \rthis{show signatures of an oscillating signal}.
They concluded that one possible explanation for the observed oscillations in the data, could be that this is a  potential signature  of non-local theories of gravity~\cite{Mazumdar,Maggiore,Frolov,Maggiore2,Shapiro,Giacchini,Netto}. Other possible reasons include statistical fluctuations of the data in one of the experiments conducted, or a periodic distance-dependent systematic feature of the data~\cite{Periv2}.

In this work, we independently assess the statistical significance of this result using multiple model selection methods motivated from Bayesian and information theoretic considerations. This complements the frequentist model comparison analysis done in P19.
We have previously used these same techniques to address a number of model selection problems in Astrophysics and Cosmology~\cite{Liu,Desai16b,Kulkarni,Ganguly,Krishak1,Krishak2}.

The outline of this manuscript is as follows. We recap the details of the \eot experiment and analysis of their data by P19 in Sect.~\ref{sec:prev}. 
We provide an abridged summary of different model comparison techniques used  in Sect.~\ref{sec:modelcomp}. We present the results of our analysis in Sect.~\ref{sec:results}. We conclude in Sect.~\ref{sec:conclusions}.
Our analysis codes are publicly available at \url{https://github.com/aditikrishak/EotWash_analysis}.

\section{Summary of \eot results and P19}
P19 independently analyzed the data from the \eot experiment~\cite{eot1}.
The \eot group  used a torsion pendulum detector to look for departures from Newton's law of gravity at sub-millimeter scales. The torsion pendulum  was suspended by a thin tungsten fiber. The gravitational interactions were measured via a torque that developed  between the holes in the torsion pendulum  and similar holes  
machined on a molybdenum detector ring below the detector. A schematic description of the \eot experiment can be found in  Fig.~1 of Ref.~\cite{eot1} or Fig.~15 of P19. The \eot group  conducted three different experiments (which they named as Experiment I, II and III), consisting of the same detector ring and  upper attractor disk,  but using different sizes for the  lower attractor disks. This feature was introduced to discriminate between potential new Physics and  systematic errors. For each of these three setups, residuals between the measured torques and the expected Newtonian  values were published. Based on these differences, 95\% c.l. upper limits were set on possible Yukawa interactions~\cite{eot1,eot2,eot3,eot4}.

P19 (and also Ref.~\cite{Periv2})   fitted the 87  residual torque data points from these experiments  ($\delta \tau$) to three different functions:
\begin{eqnarray}
\delta \tau_1(\alpha',m',r) &=& \alpha' \label{constpar} \\
\delta \tau_2(\alpha',m',r) &=& \alpha' e^{-m' r}
\label{exppar} \\
\delta \tau_3(\alpha',m',r) &=& \alpha' \cos(m' r + \phi)
\label{oscilpar}
\end{eqnarray}
These correspond to an offset Newtonian potential (which can be considered as white noise for the purpose of fitting), a Yukawa potential, and an oscillating potential respectively. Here, $\alpha'$ and $m'$ are two ad-hoc parameters, and the relations between these  parameters and the corresponding parameters of any modified gravity theory depend on the details of the experimental setup~\cite{Periv2}. \rthis{In Eq.~\ref{oscilpar}, $\phi$ fixed to $3\pi/4$ in P19, as this provided the best fit compared to other phases and all model comparison was done using this premise.}
For each of these functions, P19 carried out $\chi^2$ minimization to obtain the best-fit parameters. Among the three
models, they found that the oscillating function Eq.~\ref{oscilpar} has the smallest $\chi^2$ 
(cf. Table IV of P19), with $\Delta\chi^2 \sim -15$  with respect to the other models. They assess the statistical significance  based on this difference in $\chi^2$ with respect to the other models to be  $3\sigma$. P19 also carried out Monte-Carlo simulations of the null hypothesis and showed that the probability for the oscillation signal to be a statistical fluctuation is about 10\%. Therefore, P19 concluded from this frequentist model comparison technique that  what they have found is one of three possibilities:  either a statistical fluctuation in one of the experiments causing this anomaly, or a periodic distance-dependent systematic endemic to such laboratory-based experiments; or a signature for a short-distance modification to GR. P19 also pointed out that such an oscillating potential at short distances  is a characteristic of many non-local gravity theories~\cite{Mazumdar,Maggiore,Frolov,Maggiore2,Shapiro,Giacchini,Netto}, involving infinite derivatives  of the Lagrangian. These theories have been constructed to  solve the black hole and big-bang singularity problem~\cite{Hawking}. 

Given the potential ramification for the third possibility, we independently do a fit to the same residual torque data and assess the statistical significance of the oscillating parametrization using Bayesian and information theory-based  model selection techniques, which we 
have previously used to address multiple problems in Astrophysics and Cosmology~\cite{Liu,Desai16b,Kulkarni,Ganguly,Krishak1,Krishak2}.  This complements the frequentist analysis in P19.

\label{sec:prev}
\section{Model comparison techniques}

There are basically two distinct ways used to compare or rank two models used to fit a given dataset~\cite{Sanjib,Liddle,Liddle07}. The Bayesian analysis compares the probability of the model given the data, whereas the frequentist method compares the expected predictive accuracy of the two models for future data. The information theory techniques are a distinct class and have both frequentist and Bayesian interpretations~\cite{Liddle07}.
More details comparing and contrasting these techniques, including  the pitfalls and assumptions in  each of them can be found in Refs.~\cite{Liddle,Liddle07,Shi,Trotta,astroml,Sanjib,Weller}. A comparison of these techniques within the statistics community can be found in Ref.~\cite{Gelman}.

We briefly summarize the different selection techniques used to rank between multiple models.
More details can be found in specialized reviews~\cite{Liddle,Liddle07,Trotta,astroml,Weller,Sanjib,Gelman} or in some of  our previous works~\cite{Ganguly,Krishak1,Krishak2}. In this work we only use information theory and Bayesian techniques to compare the different models , since a  frequentist model comparison technique has already been carried out in P19.
\begin{itemize}

\item {\bf AIC and BIC:}
The Akaike Information Criterion (AIC)  as well as the  Bayesian Inference Criterion (BIC) are used to penalize for any  free parameters  to avoid overfitting. AIC is an approximate  minimization of Kullback-Leibler information entropy, which estimates the distance between two probability distributions~\citep{Liddle07}.
In this work we use a variant of AIC corrected for small sample sizes (called $\rm{AIC_c}$) and is given by~\cite{Liddle07}:
\begin{equation}
\rm{AIC_c} = -2\ln L_{max} + 2p + \frac{2p(p+1)}{N-p-1},
\label{eq:AIC}
\end{equation}
where  $p$ is the number of free parameters;
$L_{max}$ is the maximum likelihood which is also  used for parameter estimation. 

For comparing two models using AIC, the model with the lower value is the prefered model.
To assess the significance, one can apply the following ``strength of evidence'' rules for difference in AIC between two models  $\Delta \textrm{AIC} = \textrm{AIC}_i - \textrm{\
AIC}_{min}$~\cite{Shi,fabozzi}:
$\begin{array}{cc}                                  
\Delta \textrm{AIC} & \textrm{Level~of~Support~For~Model~i}\\ 
0 - 2 & \textrm{Substantial} \\                   
4 - 7 & \textrm{Considerably~Less}\\ 
> 10 & \textrm{Essentially~None} 
\end{array}$

BIC is an approximation to the Bayes factor and  is given by~\cite{Liddle}:
\begin{equation}
\rm{BIC} = -2\ln L_{max} + p \ln N, 
\label{eq:BIC}
\end{equation}
where all the parameters have the same interpretation as in Eq.~\ref{eq:AIC}. We   note that one assumption in applying BIC is that the posterior pdf is Gaussian and this may not always be satisfied. 

For the BIC, one can apply the  following ``strength of evidence'', where $\Delta \textrm{BIC} = \textrm{BIC}_i - \textrm{BIC}_{min}$~\cite{Shi,fabozzi}:

$\begin{array}{cc}                                      \Delta \textrm{BIC} & \textrm{Evidence~against~Model~i}\\                   
0 - 2 & \textrm{Not~Worth~More~Than~A~Bare~Mention}\\   2 - 6 & \textrm{Positive}\\                             6 - 10 & \textrm{Strong}\\                              >10 & \textrm{Very~Strong}                              \end{array}$

Therefore, for both AIC and BIC, the difference between two models must be greater than 10, for the model with smaller AIC/BIC to be decisively favored.

\item{\bf{WAIC:}} The Widely Applicable Information Criterion (WAIC) (also known as Watanabe-Akaike information criterion)~\cite{Watanabe} uses a Bayesian approach to quantitatively describe the ability of a model to predict new data. The model with a higher predictive power is the one with a smaller WAIC value. WAIC has some advantages compared to AIC or BIC. WAIC uses the Bayesian predictive density as opposed to AIC which uses the point estimate for the parameter~\cite{Sanjib,Ranalli,Ranalli2,Gelman}. If a model is singular, AIC and BIC do not work well, whereas WAIC works well for such cases. WAIC is also invariant under reparametrization  and in the asymptotic limit of large sample size, WAIC is equivalent to leave-one-out cross-validation~\cite{Sanjib}. 

The WAIC for a given model  is defined as~\cite{Watanabe,Ranalli,Sanjib}: 
\begin{equation}
    WAIC = -2(lppd-p_{WAIC})
    \label{eq:waic}
\end{equation}
where $lppd$ (log pointwise predictive density) is  given by~\cite{Ranalli}:
\begin{equation}
    lppd=\sum_{i=1}^N \ln \left( \frac{1}{S} \sum_{s=1}^S P(y_i|\theta^s) \right),
    \label{eq:lppd}
\end{equation}
where $S$ is the total number of posterior samples;
$P(y_i|\theta^s)$ is the likelihood for the data-point $y_i$, given the parameter vector {$\theta^s$}; and $N$ is the total number of data points. We note that in order to calculate $lppd$, one first needs to estimate the average likelihood for each data point over all the parameter samples, and then sum its logarithm over all the data points. On the other hand, in parameter estimation and also for calculating AIC/BIC, what is calculated is   the sum of log likelihood over all data points {\em for a fixed parameter set}. $p_{WAIC}$  in Eq.~\ref{eq:waic} is given by:
\begin{equation}
    p_{WAIC}= \sum_{i=1}^N \frac{1}{S-1}\sum_{s=1}^S\left(  \ln P(y_i|\theta^s) - \langle\ln P(y_i|\theta^s) \rangle \right)^2 ,
    \label{eq:pwaic}
\end{equation}
where $\langle\ln P(y_i|\theta^s) \rangle$ is the average value of $P(y_i|\theta^s)$ over all parameter samples. We note that $p_{WAIC}$ is an estimate of the effective number of free parameters in the model and can be interpreted as a penalty term, which compensates for overfitting~\cite{Ranalli,Ranalli2}. In lieu of $p_{WAIC}$, an alternate parameterization has also  been considered to compensate for the free parameters~\cite{Watanabe,Sanjib,Gelman}~\footnote{See Eq. 12 in Ref.~\cite{Gelman}}. Between the two,  Gelman et al~\cite{Gelman} have recommended the use of $p_{WAIC}$ (from Eq.~\ref{eq:pwaic}), as its series expansion is closer to leave-one out cross-validation.

Similar to AIC and BIC, the model with the smaller value of WAIC is the preferred model and the significance can be obtained using the same ``strength of evidence'' rules as used for AIC and BIC.

\item {\bf Bayesian Model Selection:}
The Bayesian model selection technique used  to compare two models ($M_1$ and $M_2$) is based on the calculation of the  Bayesian odds ratio, given by:
\begin{equation}
O_{21} = \frac{P(M_2 | D)}{P(M_1|D)},
\end{equation}
\noindent where $P(M_2 | D)$ is the posterior probability for $M_2$  given data $D$, and similarly for  $P(M_1 | D)$. 
The posterior probability for a general model $M$ is given by
\begin{equation}
P(M |D) = \frac{P(D| M) P (M)}{P (D)},
\label{eq:bi}
\end{equation}
where $P(M)$ is the prior probability for the model $M$, $P(D)$ is the probability for the data $D$ and $P(D|M)$  is the marginal likelihood or Bayesian evidence for model $M$   and  is given by:
\begin{equation}
P(D |M) = \int P (D | M,\theta) P(\theta | M) d\theta, 
\end{equation}
where $\theta$ is a vector of parameters,   encapsulating  the model $M$. If the prior probabilities of the two models are equal, the odds ratio can be written as:
\begin{equation}
B_{21} = \frac{\int P(D | M_2, \theta_2)  P(\theta_2)d \theta_2}{\int P (D | M_1, \theta_1) P (\theta_1) d\theta_1}, 
\label{eq:bayesfactor}
\end{equation}
The quantity $B_{21}$ in Eq.~\ref{eq:bayesfactor} is known as the Bayes factor. We shall  compute this quantity in order to obtain a Bayesian estimate of the statistical significance. Note that unlike the frequentist and AIC/BIC based tests, the Bayesian test does  not use the best-fit values of the parameters.

To assess the significance, a qualitative criterion based on Jeffreys' scale is used to interpret the odds ratio or Bayes factor~\cite{astroml,Trotta}. A value $> 10$ represents strong evidence in favor of $M_2$,  and a value $> 100$ represents decisive evidence~\cite{Trotta}. 
\end{itemize}
\label{sec:modelcomp}

\begin{figure*}
\centering
\includegraphics[width =\textwidth]{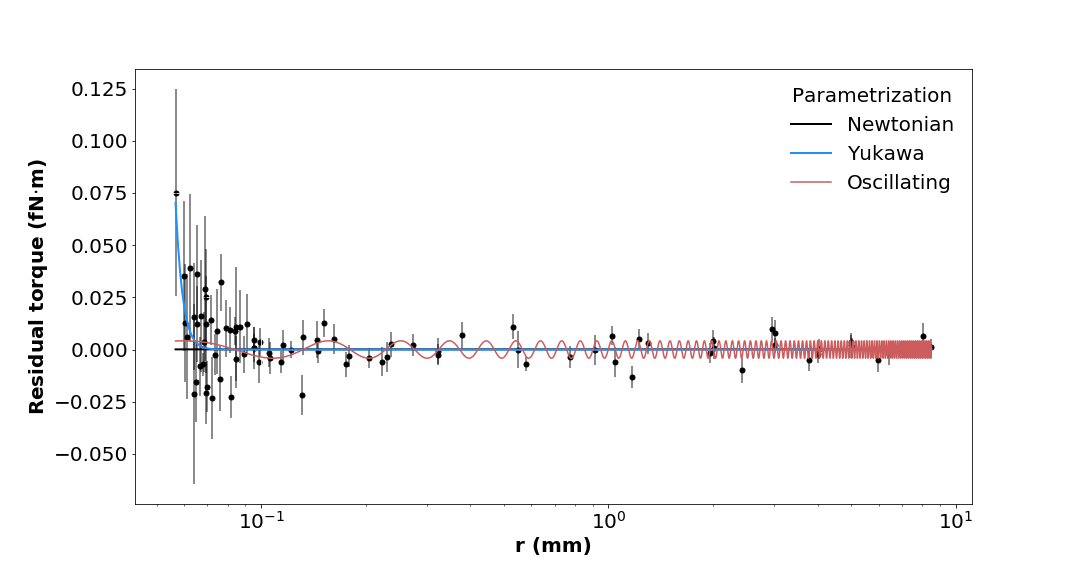}
\caption{Plots showing the data points for the residual torque (obtained from P19) from the \eot experiment, along with  the best fits obtained for the three models using $\chi^2$ minimization.
The best-fit values for each of the three models are shown in Table~\ref{table:1}. We note that for the best-fit Yukawa model, except for the first few points, the  best fit  is  indistinguishable by eye compared to the corresponding Newtonian fit.}
\label{fig1}
\end{figure*}

\begin{table}
    \centering
    
    \begin{tabular}{|c|c|c|c|}
    
    \hline
    \textbf{Parametrization}&{\textbf{$\alpha'$}} & {\textbf{$m'$}} & {\textbf{$\phi$}}\\
    {} & $\rm{(fN\cdot m)}$ & $\rm{(mm^{-1})}$ & \\
        \hline 
    Newtonian & $0.0001_{-0.0007}^{+0.0007}$ & - & -\\
    Yukawa & $(7.2 \pm 2.9) \times10^{7}$
    & $366.1^{+64}_{-47}$ 
    & - \\
    Oscillating ($\phi$ fixed) &$0.0042_{-0.0020}^{+0.0020}$ &$65.29_{-0.84}^{+0.93}$ & - \\
    Oscillating ($\phi$ free) &$0.0042_{-0.0015}^{+0.0012}$ & $65.28_{-0.21}^{+0.20}$ & $(0.756_{-0.15}^{+0.15})\pi$   \\
        
         \hline 
 \end{tabular}
\caption{The best-fit parameters obtained for the three parametrizations (Eq.~\ref{constpar}, Eq.~\ref{exppar},  Eq.~\ref{oscilpar}) by $\chi^2$ minimization. }
\label{table:1}
\end{table}

\begin{table}
    \centering
    
    \begin{tabular}{|c|c|c|c|}
    
    \hline
    \textbf{Parameter}&{\textbf{Prior}} & {\textbf{Minimum}} &  {\textbf{Maximum}}\\
        \hline 
    $\alpha'$ (Eq.~\ref{constpar}) & Uniform & -0.05 & 0.07  \\
    $\alpha'$ (Eq.~\ref{exppar}) & Uniform & 0 & $1.0 \times 10^8$  \\
    $\alpha'$ (Eq.~\ref{oscilpar}) & Uniform & 0 & 0.01  \\
    $m'$ (Yukawa) & Uniform & 0 & 500 \\
    $m'$ (Oscillating) & Uniform & 60 & 70 \\
    $\phi$ (Oscillating with phase free)& Uniform & 0 & $2\pi$ \\
     \hline 
 \end{tabular}
\caption{List of priors used to evaluate the Bayesian evidence for each of the three different models. The units for $\alpha'$  and $m'$ are in $\rm{fN\cdot m}$ and $\rm{mm^{-1}}$ respectively. }
\label{priortable}
\end{table}

\section{Analysis and Results}
\label{sec:results}

For each of the three models, we find the best-fit parameters by $\chi^2$ minimization, where $\chi^2$ is defined as follows:

\begin{equation}
\chi^2= \sum_{i=1}^N\left( \frac{\delta_{\tau_i}-\delta_{\tau_{model}}}{\sigma_{\delta_{\tau_i}}}\right)^2 ,
\label{eq:chisq}
\end{equation}
\noindent where  $\delta_{\tau_i}$ denotes the data for the residual torques (provided to us by L. Kazantzidis) and $\sigma_{\delta_{\tau_i}}$ indicates its associated error; $\delta_{\tau_{model}}$ encapsulates the three model functions (defined in Eqs~\ref{constpar},~\ref{exppar},~\ref{oscilpar}) used to fit the residual torque. We use the \rthis{\texttt{Nelder-Mead} symplex algorithm~\cite{NR} as coded in the \texttt{scipy Python} module to carry out the  optimization.}
\rthis{For the oscillating model we did two fits: with the phase fixed to $3\pi/4$ and also keeping the phase ($\phi$) as a free parameter.}
The best-fit parameters for each of these models are shown in Table~\ref{table:1} and the corresponding plots with all the three functions superposed on the data are shown in Fig.~\ref{fig1}. \rthis{We note that for the Yukawa model, we get very large values of $\alpha'$ and $m'$, compared to P19. However, our $\chi^2$ value of 82.1 for the Yukawa model is smaller than the value of  85.4 obtained in P19. }
Except for the first few points, the Yukawa best-fit is indistinguishable by eye compared to the Newtonian model. \rthis{The best-fit for $\phi$ in the oscillating parametrization we obtain is approximately the same as that in P19.} We now present results from all the different model comparison tests.

\subsection{Model Comparison using $\mathbf{AIC_c}$ and BIC}
We calculate the $\rm{AIC_c}$ and BIC values for each of the three  models in Eqs.~\ref{constpar},~\ref{exppar},~\ref{oscilpar} using equations \ref{eq:AIC} and \ref{eq:BIC}. Note that $\rm{AIC_c}$ and BIC are computed from the maximum value of the Gaussian likelihood using the best-fit parameters, obtained by $\chi^2$ minimization. These values are listed in Table~\ref{table:2}. We find that the difference in $\rm{AIC_c}$ and BIC between Yukawa model and the Newtonian one is negligible. So neither model is prefered.

However, when we compare the oscillating model with the offset Newtonian model \rthis{when the phase is fixed at $3\pi/4$}, we find that both $\Delta\rm{AIC_c}$ and $\Delta$BIC are $>10$.  Therefore, \rthis{with the phase fixed}, the oscillating model is decisively favored as compared to the Newtonian offset model. \rthis{When the phase is a free parameter, only $\rm{AIC_c}$ still shows decisive evidence for the oscillating model, whereas BIC now shows only strong evidence. This is because BIC harshly penalizes models with additional free parameters compared to AIC~\cite{Sanjib}}.

\subsection{Model Comparison using WAIC}
Using the equations \ref{eq:waic}, \ref{eq:lppd}, and \ref{eq:pwaic} we calculate the WAIC values for each of the three models with all the samples in each case.   The posterior samples for each model have been obtained by nested sampling using the \texttt{Nestle} package in \texttt{Python}. The priors used to calculate the posterior samples are the same as for Bayesian model comparison, to be described in Sect.~\ref{sec:bayesian}.

The values of WAIC obtained for the three models are listed in Table~\ref{table:2}, along with the $\Delta$WAIC values with respect to the null hypothesis (Newtonian model). We find that the results broadly agree with $\rm{AIC_c}$ and BIC. The oscillating model   has the lowest value of WAIC, implying that it has the highest predictive power among the three models.  The values of WAIC for the Newtonian and Yukawa model are almost comparable.
\rthis{When the phase in the oscillating model is fixed at $3\pi/4$}, the difference in WAIC between the Newtonian and oscillating hypothesis is greater than 10, implying that using this test also, the oscillating parametrization is decisively favored compared to the offset Newtonian model. \rthis{When the phase is kept free, the difference in WAIC falls marginally below 10 and therefore, the oscillating model is strongly favored, but not decisively.}

\subsection{Bayesian Model Comparison}
\label{sec:bayesian}
Considering the Newtonian model to be the null hypothesis, we calculate the Bayes factor for the Yukawa model   as well as  the oscillating model in comparison with the constant Newtonian  offset as the null hypothesis. We chose a Gaussian likelihood, using the data, model and experimental errors per data point. We chose uniform priors for all the parameters. A tabular summary of the priors used for each of the parametrizations can be found in Table~\ref{priortable}. We note that for the oscillating model, we have only used positive definite priors for $\alpha'$ as compared to the other two models. This is because the cosine term in the oscillating model can take both positive and negative values, so a negative value for the amplitude would be redundant. Also, as noticed in P19, there are multiple minima in ($\alpha',m'$) parameter space. In this work, we wanted to hone in on the best-fit parameter space found by P19. Therefore, for $m'$, we also used a narrow range of priors near the maximum value.  A tabular summary of the priors used for each of the parametrizations can be found in Table~\ref{priortable}.


To calculate the Bayesian evidence for each of the hypotheses, We used the \texttt{Nestle}\footnote{\url{http://kylebarbary.com/nestle/}} package in \texttt{Python}, which uses the Nested sampling algorithm~\cite{Mukherjee}.  The Bayes factor was calculated using Eq.~\ref{eq:bayesfactor} for both the Yukawa and the oscillating models in comparison with the null hypothesis. \rthis{The values for the Bayes factor are shown in Table~\ref{table:2}.}
We find that  the value for the  Bayes factor is about  29 for the Yukawa model and 544 (83) for the oscillating model, \rthis{depending on whether the phase is fixed to $3\pi/4$ or is allowed to be a free parameter} (cf. Table~\ref{table:2}).
According to the Jeffreys' scale~\cite{Trotta}, Bayesian model comparison provides ``strong'' evidence for the Yukawa model and ``decisive''/``strong'' evidence in favor of the oscillating model, \rthis{depending on whether the phase is fixed at 3$\pi$/4 or kept as  a free parameter}. \rthis{Therefore, the results from Bayesian model comparison result concur with BIC/WAIC  based information theory tests  regarding the significance of the oscillating model as compared to  the Newtonian model.}

\begin{table}
    \centering
    \begin{tabular}{|c|c|c|c|c|}
    \hline
    {} & Newtonian & Yukawa & Oscillating & Oscillating  \\
    & &  & ($\phi=3\pi/4$) & ($\phi$ free) \\
    \hline

    \textbf{$\rm{AIC_c}$ values} & -571.3 &-572.6 & -584.0 & -582.1 \\
    $\Delta \rm{AIC_c=AIC_c(Newt.) - AIC_c}$  &-&1.2&12.7 & 10.5 \\
            \hline
    \textbf{BIC values} & -568.9  & -567.6  &-579.2 & -574.7 \\
    $\Delta\rm{BIC = BIC(Newt.) - BIC} $  &-&-1.2&10.3 & 5.9 \\
        \hline
    \textbf{WAIC values} & -571.4 &-574.3& -583.6 & -579.4 \\
    $\Delta \rm{WAIC = WAIC(Newt.) - WAIC} $ & -
    & 2.9 & 12.2 & 8.0\\
    \hline
    \textbf{Bayes Factor}&- & 29 & 544 &  83\\
          \hline

 \end{tabular}
\caption{Summary of results for the model comparison tests using Bayesian and information theoretic techniques. The Bayes factor (in this table) is the ratio of Bayesian evidence for the Yukawa or oscillating model to the same for Newtonian model. $\Delta \rm{AIC_c}$ (also BIC and WAIC) indicates the difference in $\rm{AIC_c}$/BIC between the  Newtonian model and the Yukawa/oscillating model. \rthis{When the phase ($\phi$) is fixed at $3\pi/4$ similar to P19}, all the tests used here, (viz. $\rm{AIC_c}$, BIC, WAIC, and Bayes factor) decisively favor the oscillating model compared  to the Newtonian model. \rthis{When $\phi$ is a free parameter, only $\rm{AIC_c}$ shows decisive evidence for the oscillating model; whereas BIC, WAIC, and Bayes factor now only show strong evidence for the same.} }
\label{table:2}
\end{table}

\section{Conclusions}
\label{sec:conclusions}
In a series of papers, Perivolaropoulos et al~\cite{Periv2,Periv19}  independently analyzed the residual torque  data from the \eot experiment~\cite{eot1,eot2,eot3,eot4}, wherein the data was obtained after subtracting the torques  due to  a Newtonian potential. They argued that an oscillating parametrization \rthis{with a fixed phase of $3\pi/4$} provides a better fit to this data as compared to a constant term (equivalent to an offset Newtonian potential). If the oscillating fit is the true description of the data, one possible implication is that
this could be a signature of a non-local  modified theory of gravity~\cite{Mazumdar,Maggiore,Frolov,Maggiore2}.

To further investigate the viability and statistical significance of this claim, we independently analyze this same  residual torque data, and carry out a fit to the same three functions used  in Refs.~\cite{Periv2,Periv19}: offset Newtonian, Yukawa and oscillating models.
To discern the relative significance of each of the models, we carry out model comparison techniques using four different model comparison tests: $\rm{AIC_c}$ (a variant of AIC to account for small number of samples), BIC and WAIC  based information theory test, and finally a Bayesian model comparison technique based on  a  calculation of the  Bayes factor.

Our results from these model comparison analyses are summarized in Table~\ref{table:2}. 
When we compare the Yukawa and Newtonian parametrizations, we find that the difference between the models is marginal. The Bayesian test on the other hand prefers the Yukawa parametrization, where  according to Jeffreys' criterion, the support for the Yukawa model is strong, but not decisive.
However, all the four tests decisively favor the  oscillating model \rthis{with the phase fixed to the  same value as in P19. When the phase is kept as a free parameter, the oscillating model is now decisively favored using only the AIC test. Using all other tests, this model is strongly offered compared to the offset Newtonian model.}

Therefore, we agree with P19 that our complementary suite of statistical tests also favor a spatially oscillating signal in the \eot dataset, \rthis{if the phase is fixed. But once we vary the phase, the significance reduces to ``strong evidence'' using three  of the four tests used, with only AIC still showing decisive evidence.}  
However, our tests cannot discern whether this spatial oscillation is a signature of modified gravity or other systematic effects which could explain this oscillating pattern~\cite{Periv2}. Further analysis of data from other experiments is needed to  distinguish  between these possibilities.

To improve transparency in data analysis, we have made our analysis codes and datasets analyzed publicly accessible. These can be found at \url{https://github.com/aditikrishak/EotWash_analysis}.

\acknowledgments
Aditi Krishak is supported by a DST-INSPIRE fellowship. We are grateful to Lavrentios Kazantzidis for providing us the data from P19 for our analysis. We also acknowledge useful discussions with Sanjib Sharma about WAIC.

\bibliography{main}
\end{document}